\documentclass[12pt]{iopart}

\usepackage{iopams,graphicx,amssymb}
\usepackage[usenames]{color}
\usepackage[utf8x]{inputenc}
\usepackage{colortbl} 
\eqnobysec

\begin{document}

\title[Static approach to RG analysis of quenched disorder]
{Static approach to renormalization group analysis of stochastic models with spatially quenched disorder}

\author{N. V. Antonov$^1$, P. I. Kakin$^1$, N. M. Lebedev$^{1,2}$}

\address{$^1$ Department of Physics, Saint Petersburg State University,
7/9 Universitetskaya Naberezhnaya, Saint Petersburg 199034, Russia \\
$^2$ N.N. Bogoliubov Laboratory of Theoretical Physics, Joint Institute for Nuclear Research, Dubna 141980, Moscow Region, Russia}

\ead{n.antonov@spbu.ru, p.kakin@spbu.ru, nikita.m.lebedev@gmail.com}

\begin{abstract}
A new ``static'' renormalization group approach to stochastic models of fluctuating surfaces with
spatially quenched noise is proposed in which only time-independent quantities are involved. As
examples, quenched versions of the Kardar--Parisi--Zhang model and its Pavlik’s modification,
the Hwa--Kardar model of self-organized criticality, and Pastor-Satorras--Rothman model of landscape erosion are studied. It is shown that the upper critical dimension in the quenched
models is shifted by two units upwards in comparison to their counterparts with white in-time
noise. Possible scaling regimes associated with fixed points of the renormalization group equations are
found and the critical exponents are derived to the leading order of the corresponding
$\varepsilon$ expansions. Some exact values and relations for these exponents are obtained.

\end{abstract}


\section{Introduction} \label{sec:Intro}

One way to study scaling behavior of a fluctuating surface is to describe the system by stochastic differential equation for the height field $h(x)$ subjected to a random force 
$f(x)$ where $x=\{t,{\bf x}\}$ and ${\bf x}$ is the spatial coordinates at the moment $t$. Recently it was shown that the choice of this random force (or noise) can dramatically affect the scaling behavior. Precisely, the choice of the quenched disorder in the Pastor-Satorras--Rothman model of a landscape erosion \cite{Pastor1,Pastor2} proved to be the decisive factor leading to a type of scaling that explained previously thought inconclusive experimental data \cite{Delamotte}. 

Landscape erosion is a complex and multifaceted phenomenon that is quite difficult to study \cite{E1}--\cite{E14}. However, universal aspects of erosion (e.g., exponents in scaling laws)  can be studied employing relatively simple semiphenomenological models. Pastor-Satorras--Rothman model is one such model; it is continuum, nonlinear and anisotropic and was introduced in \cite{Pastor1,Pastor2} to explain the wide range of experimental measurements of the ``roughness'' exponent in the scaling law. Indeed, the exponent was found to be either small (0.30--0.55) \cite{Newman,Mark} or large (0.70--0.85) \cite{Mark}-\cite{Czi}. The authors of \cite{Pastor1,Pastor2} noticed that the large values of the exponent were observed at the small length scales and proposed that at these scales the anisotropy of the system becomes a significant factor (the flow of eroded matter goes down the slope of the landscape profile establishing a preferred direction). The model was considered with two different types of noise. 

The dynamical renormalization group (RG) analysis performed in \cite{Pastor1,Pastor2}  was later proved to have an error \cite{US}, however, the reported there seafloor and desert environment measurements confirmed that the topography along the slope is, indeed, different from the topography in the direction perpendicular to the slope. 

The further study of the Pastor-Satorras--Rothman model was performed in \cite{US}. Standard field-theoretic RG analysis revealed that the model is not renormalizable in a sense that there is an infinite number of counterterms not accounted for in the original stochastic equation. The renormalizable modification of the model with an infinite number of coupling constants was introduced. The universal relations for scaling exponents were found and it was also shown that the RG equations might have regions of infrared (IR) attractive fixed points, i.e., the model allowed for non-universal roughness exponent.

Why does existence of IR attractive regions means non-universality? Types of scaling behavior (universality classes) are differentiated by the values of scaling exponents; the exponents are related to IR attractive fixed points (because scaling laws describe asymptotic behavior in the range where the times and distances are large in comparison with characteristic microscopic scales, i.e., in the IR range). If there are several IR attractive fixed points for some values of system parameters then the system may display one of the corresponding scaling behaviors depending on the starting point. This is precisely a defining feature of non-universal scaling -- the dependence of scaling on the local parameters instead of the global ones (such as symmetries or spatial dimension, etc.).

The analysis performed in \cite{US}, however, failed to reveal explicit non-universal exponents. Moreover, only one type of noise was studied -- the``white'' (thermal or isotropic) noise, i.e., the Gaussian random noise $f(x)=f(\boldsymbol{x},t)$ with a zero mean and a prescribed pair correlation function
\begin{equation}
\langle f(x)f(x') \rangle = D_0 \,\delta(t-t')\, \delta^{(d)}({\bf x}-{\bf x}'),
\label{forceD}
\end{equation}
where $\delta(\cdot)$ is a Dirac delta function and $D_0$ is some positive amplitude. The white noise is best suited to model thermal fluctuations or external disturbances; in the case of landscape erosion it simulates the rainfalls that wash away the soil.

In the paper \cite{Delamotte} the modified model was studied non-perturbatively with functional RG \cite{NRG0,NRG1,NRG2}. It was shown that in the case of the white noise (\ref{forceD}) and spatial dimension $d=2$ there is, in fact, no regions of IR attractive fixed point but one trivial fixed point related to the universality class of ordinary diffusion (Edwards-Wilkinson fixed point \cite{EW}). However, this is not so in the case of the second type of noise. Indeed, there is an interval of IR attractive fixed points and an interval of non-universal scaling exponents.  While it is not proven that the interval of IR attractive fixed points can be reached from realistic initial conditions, the wide range of roughness exponent measurements seems to point out that non-universality is, indeed, a feature of landscape erosion at the smaller scales.

The second type of noise proposed for the model in \cite{Pastor1,Pastor2} and studied in \cite{Delamotte} is the ``columnar'' noise \cite{Caldarelli}  
\begin{equation}
\langle g(x)g(x') \rangle = D_0\, \delta^{(d)}({\bf x}-{\bf x}').
\label{forceD2}
\end{equation}

Experiments shown that heterogeneity of the soil is highly likely to be the main factor leading to scaling in erosion \cite{Czi}. This is why it is quite natural to study a model of erosion with ``quenched" (time-independent) randomness of the eroded soil, i.e., with a quenched noise:
\begin{equation}
\langle \eta(x)\eta(x') \rangle = D_0\, \Delta(h-h')\, \delta^{(d)}({\bf x}-{\bf x}'),
\label{forceD1}
\end{equation}
where $h=h(x)$ and $h'=h(x')$ are the heights of the landscape. In contrast with the correlation function for the white noise (\ref{forceD}), the correlation function (\ref{forceD1}) does not contain the delta function $\delta(t-t')$ and has an additional factor $\Delta(h-h')$. 

Such a noise also naturally appears in the problem of scaling of driven interfaces in random media near the depinning threshold where it relates to the impurities of the medium (see, e.g., \cite{Review} for a detailed review of the subject). Random interface growth with quenched noise is closely related to this problem; the so-called ``quenched'' Kardar-Parisi-Zhang stochastic differential equation is widely used as a model of the interface growth \cite{JP,KimKim}. 

However, the factor $\Delta(h-h')$ in the correlator (\ref{forceD1}) makes the task of finding an analytical solution a much more difficult one \cite{Narrayan}. Thus, non-perturbative RG and computer simulations are better suited for the study of the quenched noise (\ref{forceD1}) \cite{Wiese,Janssen2}. 

Another approach to incorporate quenched noise without foregoing an analytical solution is to use a simpler form of the noise. The columnar noise (\ref{forceD2}) is one such form that was introduced in \cite{Caldarelli} and reiterated in \cite{Pastor1, Pastor2}. Spatially quenched randomness reflects the existence of landscape regions impervious to erosion (and quenched, i.e., the regions stay non-erodible). Another term used to refer to the noise (\ref{forceD2}) is ``spatially quenched disorder'' \cite{Review}.

Spatially quenched disorder was studied in relation to the critical behavior of directed percolation in \cite{Janssen}. It was shown that the IR attractive fixed point corresponding to the scaling became unphysical after the quenched noise was used. This result was in agreement with numerical simulations reported in \cite{Moreira} where it was also shown that the scaling was described by non-universal logarithmic law. Non-universal exponents for the scaling of 1+1 dimensional directed percolation were also reported in \cite{Webman}. 

All of this and especially striking difference between results for different noises reported in \cite{Delamotte} drives the importance of considering spatially quenched noise in various stochastic models.

In this paper we apply the standard field theoretic RG to four different models of random interface growth (growth of fluctuating surfaces) with a spatially quenched noise (\ref{forceD2}) and propose a new ``static'' approach to standard RG analysis that can be used for the study of stochastic differential equations with a spatially quenched noise (\ref{forceD2}). This approach allows to forego dependence on time for all of the fields and parameters of the problem. While this approach is only slightly simpler than the standard one, non-perturbative RG or numerical simulations might benefit from the simplified form.  

We also show that the most important difference between a problem with spatially quenched noise (\ref{forceD2}) and a problem with white noise (\ref{forceD}) is their upper critical dimensions. The former has a higher upper critical dimension which may have implications for the physical interpretation and for the construction of other models. Indeed, as we show below it is precisely a higher upper critical dimension that leads to the non-trivial scaling in the model of landscape erosion with spatially quenched noise. As for the construction of other models, if one were to build a model that consists of several stochastic equations, one would need to use equations with the same upper critical dimensions. Otherwise, some of the interactions would be irrelevant for the scaling. Adding a spatially quenched noise to some of the equations may be a good solution for elevating upper critical dimension of corresponding interactions.

The plan of the paper is as follows. In section \ref{sec:Static} we discuss the spatially quenched noise and how it affects the action functional related to the stochastic equation. We also introduce the static RG approach and show how the spatially quenched noise determines the upper critical dimension.  

In sections \ref{sec:KPZ} and \ref{sec:Pavlik} we consider two isotropic models of random interface growth with spatially quenched noise (\ref{forceD2}): the Kardar-Parisi-Zhang (KPZ) stochastic differential equation \cite{KPZ} and Pavlik's modification of it proposed in \cite{Pavlik} and augmented in \cite{Pavlik2}. By employing static RG approach, it is shown that the field theory corresponding to KPZ equation is renormalizable; the fixed points of RG equations, their regions of stability and critical exponents are calculated in the leading one-loop order. It is also shown that augmented Pavlik's model has an infinite number of coupling constants and, subsequently, infinitely many RG functions. The one-loop counterterm is derived in a closed form and a two-dimensional surface of fixed points is found. If the surface contains IR attractive regions, the model exhibits scaling behavior with the non-universal critical exponents.

In sections \ref{sec:SOC} and \ref{sec:Erosion} we consider two anisotropic models of random interface growth with spatially quenched noise (\ref{forceD2}): the Hwa-Kardar continuous model of self-organized criticality \cite{Hwa} and modified  Pastor-Satorras--Rothman model of landscape erosion that involves infinite number of coupling constants \cite{Pastor1,Pastor2,Delamotte,US}. Static RG approach yields one-loop approximation for fixed points of RG equations, their regions of stability and critical exponents for the former model. For the latter model, the one-loop counterterm and a two-dimensional surface of fixed points are found. The exact relation for critical exponents differs from the one for the model with the white noise allowing for non-trivial scaling in physically interesting case of spacial dimension $d=2$. This result is compared to the one obtained in \cite{Delamotte} with non-perturbative RG.

The results and implications of the use of spatially quenched noise are further discussed in section \ref{sec:Conclusion}. 

\section{Dynamic and static approaches to RG analysis} \label{sec:Static}

Let us consider stochastic differential equation that describes behavior of some physical system
\begin{equation}
\partial_{t}h(t,\boldsymbol{x})=U(t,\boldsymbol{x},h)+f(t,\boldsymbol{x}).
\label{problem}
\end{equation}
Here $h(t,\boldsymbol{x})$ is the studied field (e.g., height of the interface profile), $U(t,\boldsymbol{x},h)$ is a given functional that depends only on the field $h$ and its spatial derivatives and, generally, consists of linear ``free'' term $Lh$ and nonlinear term $n(\phi)$, i.e., $U(h)=Lh+n(h)$. The last term in (\ref{problem}) is the Gaussian random noise $f(t,\boldsymbol{x})$ with a zero mean and the correlation function $\langle f(t,\boldsymbol{x})f(t',\boldsymbol{x'}) \rangle = D(t,\boldsymbol{x},t',\boldsymbol{x'})$. The equation (\ref{problem}) is studied on the entire $t$ axis and is supplemented by the retardation condition and by the condition that $h$ vanish asymptotically for $t\rightarrow -\infty$.

We assume that the problem (\ref{problem}) has a unique solution for any given $f$. The solution can be found by constructing a perturbation theory for the corresponding integral equation
\begin{equation}
h=\Delta_{12}[f+n(h)],
\label{pt}
\end{equation}
where $\Delta_{12}=(\partial_t - L)^{-1}$ is a retarded Green function for the linear operator $\partial_t - L$. One can use a diagrammatic representation to solve the equation (\ref{pt}) by denoting the field $h$ and the noise $f$ as tails and $\Delta_{12}$ as a line. The solution is an infinite sum of tree graphs; all of the correlation functions are represented by diagrams with lines $\Delta_{12}$ and $\Delta_{11}=\Delta_{12}D\Delta_{12}^T$ and obtained by multiplying all of the tree diagrams for all of the fields $h$ and averaging the result by $f$ (see, e.g., \cite{Book3}).

The resulting diagrammatic representation can be viewed as Feynman diagrammatic technique of a field theory with a doubled set of fields $H=\{h,h'\}$ (here $h'$ is the auxiliary field).  The linear response function $\langle\delta h(x)/\delta f(x')\rangle$ plays the role of the mixed bare propagator $\langle hh'\rangle_0$, while the other bare propagators are $\langle hh\rangle_0=\Delta_{11}$ and $\langle h'h'\rangle_0=0$.

The difference between the two diagrammatic techniques is the absence of the diagrams with closed circuits of the mixed propagators $\langle hh'\rangle_0$ in the representation of the stochastic equation. The generating functional $G(A)$ of correlation and response functions for the stochastic equation can be written in the form  \cite{MSR} (see also \cite{Book3,Zinn}):
\begin{equation}
G(A)=\int \,dh\int \,dh'\det{M} \exp{\left[ \frac{1}{2}\,h'Dh'+ h'(-\partial_t h + U(h))+AH \right]},
\label{qft2}
\end{equation}
where $A=\{a(x),a'(x)\}$ is the source, $AH=ah+a'h'$, and 
\begin{equation}
M=-\Delta_{12}^{-1}+\delta n(h)/\delta h.
\end{equation}
The factor $\det{M}$ cancels out all of the diagrams that contain close circuits of the line $\Delta_{12}=\langle hh'\rangle_0$. However, $\Delta_{12}$ is a retarded Green function which means that all those diagrams vanish. This allows one to drop $\det{M}$ from the expression (\ref{qft2}). Thus, any stochastic problem (\ref{problem}) is equal to the field theory with a double number of fields $H=\{h,h'\}$ and action functional $S(H)$:
\begin{equation}
S(H)=\int dt\int d{\bf x} \,\, \left(\frac{1}{2}\,h'Dh'+h'\left\{-\partial_{t}h+U(h)\right\}\right).
\label{act}
\end{equation}
In other words, correlation functions and response functions of the problem (\ref{problem}) are identified with various Green functions of the field theory with the action functional (\ref{act}) and defined by generating functional
\begin{equation}
G(A)=\int dh \int dh' \exp{\left[ S(H)+AH\right]}.
\label{gf}
\end{equation} 

This ``dynamic'' approach holds true both for the white noise (\ref{forceD}) and for the spatially quenched disorder (\ref{forceD2}) (to the models studied in the present paper it was applied, e.g., in \cite{Delamotte,Vestnik}). For the latter case, though, another (static) approach is possible. Before discussing it further, let us first list the main differences in standard approach between the problems with the white noise (\ref{forceD}) and with the quenched noise (\ref{forceD2}).

The first term in (\ref{act}) for the case of white noise (\ref{forceD}) is
\begin{equation}
\int dt\int d{\bf x} \,\, \frac{1}{2}\,h'D_0 h'.
\label{1t1}
\end{equation}
However, the quenched noise (\ref{forceD2}) does not contain the delta function $\delta(t-t')$. Thus, the first term has additional integration over time:
\begin{equation}
\int dt\int dt'\int d{\bf x} \,\, \frac{1}{2}\,h'D_0 h'.
\label{1t2}
\end{equation}
This difference turns out to play quite an important role. To illustrate that, let us recall the connection between the  ultraviolet (UV) divergences and the canonical dimensions of the parameters and fields of the problem (\ref{problem}) \cite{Book3,Zinn}. If $T$ is the time scale  and $L$ is the length scale, then canonical dimension of some quantity $F$ is described by the
frequency dimension $d_{F}^{\omega}$ and the momentum dimension $d_{F}^{k}$:
\[[F] \sim [T]^{-d_{F}^{\omega}} [L]^{-d_{F}^{k}}.\]
The normalization conditions
\[ d_k^k=-d_{\bf x}^k=1,\ d_k^{\omega} =d_{\bf x}^{\omega }=0,\
d_{\omega }^k=d_t^k=0,\  d_{\omega }^{\omega }=-d_t^{\omega }=1 \] are assumed; each term of the action functional must be dimensionless. The total canonical dimension is defined as $d_{F}=d_{F}^{k}+2d_{F}^{\omega}$
(in the free theory, $\partial_{t}\propto\boldsymbol{\partial}^{2}$). 

The UV divergences in the Green functions manifest themselves as poles in $\varepsilon=d_{*}-d$, where $d_{*}$ is the upper critical dimension (i.e., the spatial dimension at which all of the coupling constants simultaneously become dimensionless and corresponding nonlinear terms in action functional become marginal in the sense of Wilson). 

The addition of the integration $\int dt'$ in (\ref{1t2}) changes the total canonical dimension of the term $h'D_0 h'$. If it is equal to $d_{h'D_0 h'}$ in the case of the the white noise (\ref{forceD}), then it will be equal to $2+d_{h'D_0 h'}$ in the case of the quenched noise (\ref{forceD2}). Thus, the upper critical dimension will be shifted upwards by two as well. All of the nonlinear terms in $n(h)$ that are irrelevant at the new upper critical dimension will no longer affect IR asymptotic behavior of the studied system. 

Thus, the choice of the quenched noise (\ref{forceD2}) strongly affects the IR behavior of the model.

Another difference between the choices of the noises is the propagator $\Delta_{11}=\langle hh \rangle_0$. If it is the propagator for the white noise (\ref{forceD}) then the propagator ${\tilde \Delta_{11}}$ for the quenched noise (\ref{forceD2}) is a subject to the following relations:
\begin{equation}
\int dt'\,\, \Delta_{11}(t',\boldsymbol{x})={\tilde \Delta_{11}}(\boldsymbol{x}), \quad {\tilde \Delta_{11}}(\omega,\boldsymbol{k})=2\pi\delta(\omega)\Delta_{11}(\omega,\boldsymbol{k}).
\label{propnew}
\end{equation}

The divergence index of an arbitrary 1-irreducible Green function $\Gamma$ is the total dimension $\delta_{\Gamma}=d+2-d_h N_h-d_{h'}N_{h'}|_{d=d^{*}}$ taken at the upper critical dimension $d_{*}$. Here $N_h,N_{h'}$ are the numbers of corresponding fields entering
into the function $\Gamma$. Superficial UV divergences can only be present in the 1-irreducible functions that correspond to the non-negative index of divergence $\delta_{\Gamma}$ (see, e.g., \cite{Book3}). However, in the case of diagrams with $n$ ($n>1$) inner lines ${\tilde \Delta_{11}}$, the real divergence index is given by $\delta_{\Gamma}'=\delta_{\Gamma}+2(n-1)$. Indeed, such a diagram  will have $(n-1)$ delta functions of external frequency as factors and, since $d_{\delta(\omega)}=-2$, there will be an increase of the divergence index by $2(n-1)$.

Galilean symmetry often plays an important role in the analysis of stochastic models like (\ref{problem}). It should be noted that inclusion of the time-independent noise (\ref{forceD2}) destroys that symmetry even if it was present in the model with the white noise (\ref{forceD}).

Now let us consider an alternative approach (referred as static approach from now on) to the problem (\ref{problem}) with the spatially quenched disorder (\ref{forceD2}). Let us treat the noise as time-independent, i.e., $g=g(\boldsymbol{x})$. Then it is natural to look for the solution in the form $h(\boldsymbol{x})$, too. The equation (\ref{problem}) becomes
\begin{equation}
U(\boldsymbol{x},h)+g(\boldsymbol{x})=0.
\label{problem2}
\end{equation}

The perturbative diagrammatic construction sketched below equation (\ref{pt}) equally applies to this case. Within the perturbation theory, the solution of (\ref{problem2}) exists and is unique. The analog of representation (\ref{qft2}) takes on the form

\begin{equation}
G(A)=\int \,dh\int \,dh'\det{M} \exp{\left[ \frac{1}{2}\,h'Dh'+ h'U(h)+AH \right]}.
\label{qft3}
\end{equation}

As before, the role of the determinant $\det{M}$ is the cancellation of the superfluous diagrams with closed circuits of the mixed propagators $\Delta_{12}$. They are all nontrivial now, and the determinant $\det{M}$ contains nontrivial dependence on the field $h$. 

In analogy with the gauge theories, we could represent
the factor $\det{M}$ as a functional integral over fermion ghost fields. There, the diagrams with circuits of the ghost propagators partly cancel the diagrams with circuits of gauge fields propagators to ensure causality. Then, the global BRST symmetry ensures that the structure of the fermion-boson action functional is preserved under renormalization (see, e.g., \cite{Zinn}, Ch. 16 and 17). 

In our case, however, the situation is simpler because the superfluous diagrams are cancelled by the circuits of the ghost fields exactly. Thus, one can simply consider the field theory with the static action 
\begin{equation}
S(H)=\int d{\bf x} \,\, \left(\frac{1}{2}\,h'Dh'+h'U(h)\right)
\label{act2}
\end{equation}
and simultaneously omit the superfluous diagrams and the determinant $\det{M}$.

The generating functional (\ref{gf}) gives correlation functions and response functions of the problem (\ref{problem2}) if one compliments the theory (\ref{act2}) with restriction that diagrams that contain closed circuits of the line $\Delta_{12}$ vanish. This procedure commutes with the renormalization procedure (R-operation and the minimal subtraction (MS) scheme). 

One can  check directly from the diagrams that the equal-time correlation functions of the dynamic model (\ref{act}) are equal to the (time-independent) correlation functions of the static model (\ref{act2}). The relation between the response functions is a bit more tricky. Comparison of the expression (\ref{1t2}) and the first term in (\ref{act2}) shows that the auxiliary fields in the two models are related as
\begin{equation}
h'({\bf x} ) = \int dt \, h' (t,{\bf x}).
\end{equation}
Thus, the response function $G({\bf x};{\bf x'}) = \langle h'({\bf x}) h({\bf x'}) \rangle$  of the static model and its dynamic counterpart $G(t, {\bf x};t',{\bf x'}) = \langle h'(t,{\bf x}) h(t',{\bf x'}) \rangle$  are related via the integration over the time variable:
 \begin{equation}
G({\bf x};{\bf x'})= \int \, dt\, G(t, {\bf x};t',{\bf x'}).                       \label{***}
 \end{equation}
The latter is capable to describe the response of the solution to a time-dependent perturbation while the former deals only with the time-independent disturbances.

In this static approach to renormalization, the canonical frequency dimension of every field and parameter is zero. One of the parameters of the static problem (\ref{problem2}) can be scaled out; then, the canonical momentum dimensions are the same as the ones in the dynamic approach for any field and parameter except for $h'$ and $D_0$. 

The static approach is simpler than the dynamic approach. In particular, it seems likely that non-perturbative RG may benefit from its use.

\section{Static approach to RG analysis of the KPZ model} \label{sec:KPZ}

The KPZ nonlinear stochastic differential equation was proposed in \cite{KPZ} to describe universal properties of kinetic roughening of randomly growing surfaces (including phase boundaries):
\begin{equation}
\partial_{t} h= \nu_0\, \partial^{2} h +
\lambda_{0}(\partial h)^{2}/2 + g.
\label{KPZ0}
\end{equation}
Here $h(x)=h(t,{\bf x})$ is the height of the surface profile, 
$\partial_{t}= \partial/\partial t$,
$\boldsymbol{\partial}=\{\partial_{i}\}= \{\partial/\partial x_{i}\}$, 
$\boldsymbol{\partial}^{2}=(\boldsymbol{\partial}\cdot\boldsymbol{\partial})=\partial_{i}\partial_{i}$  and
$(\boldsymbol{\partial} h)^{2}=(\boldsymbol{\partial} h\cdot\boldsymbol{\partial}h) = \partial_{i}h\partial_{i}h$ (the summations over
repeated tensor indices are implied throughout the paper). The first term in (\ref{KPZ0}) describes the surface tension while the second term stands for the excess growth along the local normal to the surface. The last term is the random noise $g=g(x)$ chosen here in the form (\ref{forceD2}). The coefficient $\lambda_{0}$ can be dropped from the equation (\ref{KPZ0}) (it can be absorbed by the fields and other parameters of the model) so we set $\lambda_{0}=1$ in the following.

Kinetic roughening can be observed in a wide range of physical systems: in flame and smoke propagation, in the growth of colloid aggregates and tumours, in a deposition of a substance on a substrate, and so on~\cite{rost1}--\cite{ball}. A randomly kinetically growing surface becomes rougher and rougher with time in a sense that IR  asymptotic behavior of the surface is described by a power law with universal exponents \cite{rost1}--\cite{rost3} 
\begin{equation}
\langle\left[h(t,{\bf x}) - h(0,{\bf 0})\right]^{2}\rangle \simeq
r^{2\chi}\, F (r/t^{1/z}), \quad r=|{\bf x}|.
\label{scaling}
\end{equation}
Here the brackets $\langle\dots\rangle$ denote averaging over the statistical ensemble, $\chi$ and $z$ are respectively the roughness exponent and the dynamical exponent that depend only on the global parameters of the system, and $F$ is a certain universal scaling function. 

While there is a number of microscopic models describing the kinetic roughening (\cite{EW,Eden,SOS,ball}, etc.), its universal aspects can be found by RG analysis of simplified models for a smoothed (coarse-grained) height field such as the KPZ model (\ref{KPZ0}). The KPZ model is a $d$-dimensional generalization of the Burgers equation and can be
mapped onto a model of directed polymers in random media and on a model of Bose many-particle system with attraction; see e.g. \cite{Burg}. A number of modifications of KPZ model were introduced in \cite{KimKim,Color}--\cite{Us-an}. 

The KPZ model is miltiplicatively renormalizable \cite{KPZ,FNS,10,11} and its upper critical dimension is $d_{*}=2$ when the random noise is white one (\ref{forceD}).
The nontrivial fixed point of the RG equations corresponds to the universality class with the exponents $\chi=0$, $z=2$ and becomes IR attractive when $2-d>0$ (d is a spatial dimension); its coordinates, however, lie outside the physical range of the model parameters. Nevertheless, non-perturbative RG is able to find another ``essentially non-perturbative'' IR attractive fixed point \cite{Canet,Canet2} that is physically acceptable (but cannot be detected within any kind of perturbative analysis).

The stochastic problem
(\ref{KPZ0}) is equivalent to the field theory
with the set of fields $H=\{h,h'\}$ and the action functional 
\begin{equation}
S(H)= \int d\boldsymbol{x} \,\,\left\{\frac{1}{2}\,h' D_0 h'+h'\partial^{2} h+\frac{1}{2}h'(\partial h)^2\right\}.
\label{act3}
\end{equation}
Here we set $\nu_0=1$ because it can be absorbed by other fields and parameters of the system.

The bare propagators are as follows:
\begin{equation}
\langle hh' \rangle_{0} = \langle h'h \rangle_{0}^{*} = \left\{-i \omega+  k^2 \right\}^{-1}, \quad
\langle hh \rangle_{0} =  2\pi \delta(\omega)D_0/k^4.
\label{prop1}
\end{equation}

Canonical dimensions of the fields and the parameters of the theory (\ref{act3}) are stated in the table~\ref{tab1}. The coupling constant $g_0=D_0$ is dimensionless when $d=4$, thus, $d_{*}=4$ is the upper critical dimension and UV divergences in the Green functions are poles in $\varepsilon=4-d$. If $\Lambda$ is a typical UV momentum scale of the problem then $g_0\sim {\Lambda}^{\varepsilon}$. The table~\ref{tab1} also contains renormalized coupling constant $g$ and the renormalization mass $\mu$ (defined by its canonical dimension).

\begin{table}[h]
  \begin{center}
   \small
    \begin{tabular}{|c|c|c|c|c|c|c|}
     \hline
     $F$ & $h'$ & $h$  & $g_{0}$ & $g$ & $\mu$\\
     \hline

$d_{F}$ & $d-2$ & $0$  & $4-d$ & $0$ & $1$\\
     \hline
    \end{tabular}

  \end{center}
 \caption{\label{tab1}Canonical dimensions of the fields and the parameters of the theory (\ref{act3}).}
\end{table}

The index of the UV divergence for of an arbitrary 1-irreducible Green function $\Gamma$ in the theory (\ref{act3}) is 
\begin{equation}
\delta_{\Gamma}'= \delta_{\Gamma} - N_{h}=4 - N_{h} - 2N_{h'}.
\label{IndeX}
\end{equation}
Here $N_h,N_{h'}$ are the numbers of fields entering
into the Green function $\Gamma$. The formal index $\delta_{\Gamma}$ was adjusted by $N_{h}$ because the field $h$ appears in the vertex $h'(\boldsymbol{\partial} h)^{2}$ only under spatial derivative (its appearance in the Green function $\Gamma$ gives an external momentum that should be accounted for in the real divergence index $\delta_{\Gamma}'$).

Analysis of the divergence index shows that the theory (\ref{act3}) is multiplicatively renormalizable because superficial UV divergences can be present only in the
1-irreducible functions $\langle h'hh \rangle_{1-\mbox{{\small ir}}} $, 
$\langle h'h \rangle_{1-\mbox{{\small ir}}}$, $\langle  h'h' \rangle_{1-\mbox{{\small ir}}}$. 

The renormalized action is given by the action (\ref{act3}) with the fields and the parameters being replaced with their renormalized counterparts: $h\rightarrow Z_h h$, $h'\rightarrow Z_{h'} h'$, $D_{0}=g_{0} \rightarrow D=Z_{g}g\mu^{\varepsilon}$:
\begin{eqnarray}
S_{R} (H)=\int d\boldsymbol{x} \,\, \left\{\frac{1}{2}Z_1 h'Dh'+Z_{2}h'\partial^{2}h +  \frac{1}{2}Z_3 h'(\partial h)^2\right\}.
\label{RenAct3}
\end{eqnarray}
The renormalization constants are related as follows:
\begin{equation}
Z_{g}=Z_{1}Z_{2}^{-4}Z_{3}^{2}, \quad Z_{h}=Z_{3}Z_{2}^{-1}, \quad Z_{h'}=Z_{3}^{-1}Z_{2}^{2}.
\label{Zg3}
\end{equation}
The renormalization constants $Z_{1}$, $Z_{2}$ and $Z_{3}$ are calculated to the first order in $g$ (one-loop approximation) directly from the diagrams by utilizing the minimal subtraction (MS) scheme:
\begin{eqnarray}
Z_{1}= 1- \frac{{\tilde g}}{2\varepsilon}, \quad 
Z_{2}=Z_{3}= 1+ \frac{{\tilde g}}{4\varepsilon}.
\label{Zoneloop3}
\end{eqnarray}
Here ${\tilde g}=g S_d/(2\pi)^d$ and $S_{d}=2\pi^{d/2}/\Gamma(d/2)$ is the area of the unit sphere in $d$ dimensions. 

RG equations for the renormalized Green functions $G_{R}(g,\mu,\dots)$ (the ellipsis stands for the times,
coordinates, etc.) are as follows (see, e.g., \cite{Book3,Zinn}):
\begin{equation}
\left\{ D_{RG} + N_{h} \gamma_{h} + N_{h'} \gamma_{h'} \right\}\,G_{R}(e,\mu,\dots) = 0,
\label{RG13}
\end{equation}
where $D_{RG}$ is the differential operation:
\begin{equation}
D_{RG}\equiv D_{\mu} + \beta_{g}\partial_{g},
\label{RG23}
\end{equation}
and the anomalous dimensions $\gamma_{h},\gamma_{h'}$ and the $\beta$-function are defined as follows
\begin{equation}
\gamma_{F}\equiv \widetilde D_{\mu} \ln Z_{F}, \quad \beta_{g}\equiv \widetilde D_{\mu}g =g(-\gamma_{g}-\varepsilon).
\label{RGF13}
\end{equation}
Here $D_{x}\equiv x\partial_{x}$ for any variable
$x$ and $\widetilde D_{\mu}$ is operation $D_{\mu}$ with the fixed bare parameters. 

Equations (\ref{Zg3}),(\ref{Zoneloop3}),(\ref{RGF13}) give in the one-loop approximation:
\begin{eqnarray}
\gamma_{g}= {\tilde g}/2, \quad \beta_g=-g({\tilde g}/2+\varepsilon).
\label{an1}
\end{eqnarray}

IR asymptotic behavior of 1-irreducible Green functions is related to the IR attractive fixed points of RG equations. The coordinates of the fixed points are found from the following equation:
\begin{equation}
\beta (g_{*}) =0.
\label{fp3}
\end{equation}
IR attractive fixed point is the point for which the derivative $\partial\beta/\partial g$ is positive. In the theory (\ref{RenAct3}) there are two fixed points:
\begin{enumerate}
\item the Gaussian (free) fixed point:
\begin{eqnarray}
g^{*}=0, \quad \partial\beta/\partial g|_{g^{*}}= - \varepsilon.
\end{eqnarray}
\item the fixed point:
\begin{eqnarray}
g^{*}=-2\varepsilon, \quad \partial\beta/\partial g|_{g^{*}}= \varepsilon.
\label{fpKPZ}
\end{eqnarray}
\end{enumerate}
The fixed point (\ref{fpKPZ}) is IR attractive when $d=2$ (the most interesting dimension from the physical standpoint) but it has $g^{*}<0$ (i.e. $D<0$). Thus, the IR attractive fixed point of the problem (\ref{KPZ0}) with spatially quenched disorder (\ref{forceD2}) lies in unphysical region. 

The critical dimension $\Delta_{F}$ of a quantity $F$ is given by the following expression (normalization condition $\Delta_{k} = 1$ is assumed):
\begin{eqnarray}
\Delta_{F} = d^{k}_{F} + \gamma_{F}^{*},
\label{dim1}
\end{eqnarray}
where $d^{k}_{F}=d_{F}$ is the the canonical
dimension of $F$ from the table~~\ref{tab1} and $\gamma_{F}^{*}$ is the anomalous dimension taken at the fixed point (see, e.g., \cite{Book3,Zinn}).

We obtain for the fields $h,h'$:
\[ \Delta_{h} =  \gamma_{h}^{*}, \quad
\Delta_{h'} = d-2 + \gamma_{h'}^{*}. \]
Critical dimensions in one-loop approximation are
\begin{eqnarray}
\Delta_{h} = 0, \quad \Delta_{h'} = d-2
\label{dimD}
\end{eqnarray}
for the Gaussian fixed point and
\begin{eqnarray}
\Delta_{h} = 0, \quad \Delta_{h'} = d/2
\label{dimD2}
\end{eqnarray}
for the fixed point (\ref{fpKPZ}). To relate the critical dimensions with the critical exponents from (\ref{scaling}) one has to identify $\Delta_{h}= - \chi$ and $\Delta_{\omega}=z$.

The KPZ equation with the spatially quenched noise has a higher upper critical dimension than the original equation. Thus, the nontrivial fixed point remains IR attractive at $d=2$ while it became trivial at $d=2$ in the original problem \cite{KPZ,FNS,10,11}. However, this fixed point is unphysical like its counterpart in the original KPZ model for $d<2$. 

\section{Static approach to RG analysis of Pavlik's model} \label{sec:Pavlik}
The KPZ equation (\ref{KPZ0}) was modified in \cite{Pavlik}:
\begin{equation}
\partial_{t} h= \nu_0\, \partial^{2} h + \partial^2 h^{2}/2 + f.
\label{KPZ24}
\end{equation}
Here $f$ is the white noise (\ref{forceD}). However, the proposed model (Pavlik's model) was proved to be non-renormalizable in \cite{Pavlik2}. It was shown that the model becomes renormalizable if it is modified to include the whole series in the powers of $h$:
\begin{equation}
\partial_{t} h= \nu_0\, \partial^{2} h + \partial^{2} V(h)+ f,
\label{KPZ34}
\end{equation}
where function $V(h)$ is the following series:
\begin{equation}
V(h)=\sum_{n=2}^{\infty}\frac{\lambda_{n0} h^{n}}{n!}.
\label{Vh4}
\end{equation}
Let us study the model (\ref{KPZ34}) with a spatially quenched disorder (\ref{forceD2}) using the method of analysing models with infinite number of coupling constants suggested in \cite{Pavlik2,AA}.

The problem (\ref{KPZ34}) is equivalent to the field theory with the set of fields  $H=\{h,h'\}$ and action functional:
\begin{equation}
S(H)=\int d{\bf x} \,\, \left\{ h'h'+h'\partial^{2} h +
h'\partial^{2}\sum_{n=2}^{\infty}\frac{\lambda_{n0} h^{n}}{n!}\right\}.
\label{act4}
\end{equation}
Here we set $D_0=2$ in (\ref{forceD2}) and $\nu_0=1$ as we did in the previous Section.
\begin{table}[h]
  \begin{center}
   \small
    \begin{tabular}{|c|c|c|c|c|c|c|c|}
     \hline
     $F$ & $h'$ & $h$ &   $\lambda_{n0}$ & $g_n$ & $\mu$\\
     \hline
$d_{F}$ & $d/2$ & $d/2-2$ & $(1-n)(d-4)/2$ & $0$ & $1$\\
     \hline
    \end{tabular}

  \end{center}
 \caption{\label{tab24}Canonical dimensions of the fields and the parameters of the theory (\ref{act4}).}
\end{table}

The canonical dimensions of the fields and the parameters of the theory (\ref{act4}) are stated in the table~\ref{tab24}. The upper critical dimension is $d_{*}=4$. The real index of the UV divergence is 
\begin{equation}
\delta_{\Gamma}'= 4 - 4N_{h'}.
\label{IndeX2}
\end{equation}
The superficial UV divergences can be present only in the
1-irreducible functions of the form $\langle  h'h\dots h \rangle_{1-ir} $ with 
the counter-term $(\partial^2 h')h^n$ (for any $n\geq 1$). Thus, the theory (\ref{act4}) is multiplicatively renormalizable and involves an infinite number of coupling constants:
\begin{equation}
S_{R} (H)=\int d\boldsymbol{x} \,\, \left\{h'h'+Z_{1}h'\partial^{2}h + h'\partial^{2}\sum_{n=2}^{\infty}\frac{Z_n\lambda_{n} h^{n}}{n!}\right\}.
\label{RenAct34}
\end{equation}
The renormalized counterparts for the coupling constants $g_{n0}=\lambda_{n0}$ are $\lambda_{n}=\mu^{\varepsilon(n-1)/2} g_{n}$
where $\varepsilon=4-d$ and $\mu$ is the renormalization mass. There is an infinite number of $\beta$-functions:
\begin{equation}
\beta_{n} = g_n\,[-\varepsilon(n-1)/2-\gamma_{g_n}],
\label{betagw24}
\end{equation}
where $\gamma_{g_n}$ are anomalous dimensions for the coupling constants $g_n$. 

Let us consider expansion in the number $p$ of loops of the generating functional $\Gamma_{R}(H)$ of the 1-irreducible Green functions of the theory (\ref{RenAct34}):
\begin{equation} 
\Gamma_{R}(H)=\sum_{p=0}^{\infty}  \Gamma^{(p)}(H),\
\Gamma^{(0)}(H) = S_{R} (H).
\label{W194}
\end{equation}
The loopless (tree-like) contribution $\Gamma^{(0)}(H)$ is the action $S_{R} (H)$ while the one-loop contribution can be calculated using the functional analysis method suggested in \cite{Pavlik2,AA}. Here we omit the details for brevity; the analysis needed is quite straightforward. Thus, the divergent part of $\Gamma^{(1)}(H)$ in one-loop approximation is expressed as follows:
\begin{equation}
\Gamma^{(1)}(H)= \frac{S_d}{(2\pi)^d}\frac{\mu^{-\varepsilon}}{ \varepsilon}\int d{\bf x}
\frac{V''(h({\bf x}))}{ (V'(h({\bf x})))^2 }\, \partial^{2}h'({\bf x})
\label{I184}
\end{equation}
Here function $V$ is treated as a function of a single variable $h({\bf x})$, and $V'$, $V''$ as the corresponding derivatives with respect to this variable.

The poles in $\varepsilon$ in the sum of (\ref{I184}) and the loopless contribution in (\ref{W194}) cancel each other out. One can use it to find the one-loop contributions of order $1/\varepsilon$ in renormalization constants $Z$.

Let us introduce the representation
\begin{equation}
\frac{V''(h(x))}{ (V'(h(x)))^2 } =\sum^{\infty}_{n=0}
\mu^{\varepsilon(n+1)/2} \frac{r_{n}h^{n}}{n!}.
\label{II194}
\end{equation}
Then $r_{n}$ are completely dimensionless coefficients and polynomials in the coupling constants $g_n$. Now we can obtain renormalization constants in one-loop approximation:
\begin{equation}
Z_{1}=1-\frac{r_{1}S_d}{(2\pi)^d\varepsilon}, \quad
Z_{n}=1-\frac{r_{n}}{g_{n}}\frac{S_d}{(2\pi)^d\varepsilon}.
\label{I204}
\end{equation}
The relations between renormalization constant  $Z_{h}=Z_{1}$, $Z_{g_{n}}=Z_{n}Z_{1}^{-n}$, expression (\ref{betagw24}) and definition of anomalous dimensions gives the following expressions for the one-loop RG-functions:
\begin{equation}
\gamma_{h}=a D_{g} r_{1}/2;
\label{I22a4}
\end{equation}
\begin{equation}
\beta_{n}=-\varepsilon\frac{n-1}{2}g_{n}+ng_{n}\gamma_{h}-\frac{a}{2}(D_{g}-n+1) r_{n},
\label{I22b4}
\end{equation}
where $a\equiv 2S_{d}/(2\pi)^d$ and $D_{g}\equiv \sum^{\infty}_{n=2} (n-1) g_{n} \partial_{g_{n}}$.

Let us consider the explicit expressions for the first four coefficients $r_n$ (for internal consistency
we suppose that $g_n \simeq g_2^{n-2}$):

\[ r_{1}=g_{3}-2g_{2}^{2},\quad r_{2}=g_{4}-6g_{2}g_{3}+6g_{2}^{3}, \]
\[ r_{3}=g_{5}-8g_{2}g_{4} -6g_{3}^{2}+36g_{2}^{2}g_{3}-24g_{2}^{4}, \]
\[  r_{4}=g_{6}-10g_{2}g_{5} +60g_{2}^{2} g_{4} -20 g_{3}g_{4}+
90 g_{2}g_{3}^{2}-240g_{2}^{3}g_{3} +120g_{2}^{5} , \]
when substituted into (\ref{I22a4}), (\ref{I22b4}) they yield: 
\begin{eqnarray}
\gamma_{h} &=& a(g_{3}-2g_{2}^{2}),
\label{II23a4} \\
\beta_{2}&=&-\frac{\varepsilon}{2} g_{2}+a(-g_{4} +8 g_{2}g_{3}-10 g_{2}^{3}),
\nonumber \\
\beta_{3}&=&-\varepsilon g_{3}+a(-g_{5}+8g_{2}g_{4}+9 g_{3}^{2}-42 
g_{2}^{2}g_{3} +24 g_{2}^{4}),
\nonumber \\
\beta_{4}&=&-\frac{3}{2}\varepsilon g_{4}+a(-g_{6}+10g_{5}g_{2}+24g_{4}g_{3}-68g_{4}g_{2}^{2}+240g_{3}g_{2}^{3}-\nonumber\\
&&90g_{3}^{3}g_{2}-120g_{2}^{5}).
\label{II23b4}
\end{eqnarray}
These examples ($\beta_{2}$, $\beta_{3}$, $\beta_{4}$) illustrate the general form of the $\beta$-functions (\ref{I22b4}).

When finding fixed point of RG equations, one can choose coordinates $g_{2*}$ and $g_{3*}$ arbitrarily, then the other coordinates $g_{n*}$ with $n\ge4$ will be uniquely determined through equations $\beta_{k}(g_{*})=0$, $k\ge2$. Thus, there is a two-dimensional surface of fixed points parametrized by the values of $g_{2*}$ and $g_{3*}$ in the infinite-dimensional space of the couplings $g\equiv \{ g_{n} \}$. 

If this surface of fixed points contains regions of IR attractive fixed points, then the theory (\ref{RenAct34}) predicts IR scaling with 
non-universal critical dimensions. The critical dimensions are non-universal in the sense that they depend on the the parameters $g_{2*}$ and $g_{3*}$, and, therefore, on the initial values of all the coupling constants.

Lastly, expression (\ref{II23a4}) yields one-loop approximation for the critical dimension $\Delta_{h}$ of the field $h$: $\Delta_{h}=d/2-2+a (g_{3*}-2g_{2*}^{2})$.

\section{Static approach to RG analysis of the Hwa-Kardar model} \label{sec:SOC}

Various equilibrium systems display critical scaling behavior near their second-order phase transition
points (see, e.g., \cite{Book3,Zinn}). Open nonequilibrium systems with dissipative transport may display critical scaling too despite not including a tuning parameter
(like the temperature). This phenomenon of achieving the critical state as a result of system's intrinsic dynamics is called self-organized criticality (SOC) and is believed to be ubiquitous in the Nature with various examples found in biological, ecological and social systems \cite{32}-\cite{Bak}. 

SOC is usually described by discrete models with discrete time steps; here, however, we would like to consider anisotropic  continuous model proposed in \cite{Hwa} and written for a smoothed (coarse-grained) height field $h(x)$. This model with the effects of turbulent advection taken into account was studied in \cite{AK1} while it was also discussed in connection with landscape erosion in \cite{Pastor1,Pastor2,Tadic}. 

Let us first introduce anisotropy. Let ${\bf n}$ be a unit constant vector that determines a certain preferred
direction (in connection to the interface models it is usually the direction down the slope of the interface profile). Then any vector ${\bf x}$ can be expressed as
${\bf x} = {\bf x}_{\bot} + {\bf n} x_{\parallel}$ where
${\bf x}_{\bot} \cdot {\bf n} =0$. Let
$\partial_{i} = \partial/ \partial {x_i}$ with $i=1\dots d$
be the derivative in the full $d$-dimensional ${\bf x}$ space, then
$\partial_{\bot}=\partial/ \partial {x_{\bot i}}$ with $i=1\dots d-1$
is the derivative in the subspace orthogonal to ${\bf n}$, and
$\partial_{\parallel} = {\bf n}  \cdot \partial$.

The stochastic differential equation describing the height field
$h(x)=h(t,{\bf x})$ of the system with SOC is \cite{Hwa}:
\begin{equation}
\partial_{t} h= \nu_{\bot 0}\, \partial_{\bot}^{2} h + \nu_{\parallel 0}\,
\partial_{\parallel}^{2} h -
\partial_{\parallel} h^2/2 + g,
\label{soc}
\end{equation}
where  $\nu_{\bot 0}$, $\nu_{\parallel 0}$ are viscosity coefficients. We choose $g$ to be a spatially quenched noise (\ref{forceD2}) instead of the white one considered in \cite{Hwa}.

The problem (\ref{soc}) is equivalent to the
field theory of the set of fields $H=\{h,h'\}$ with action functional:
\begin{equation}
S(H)=\int d{\bf x} \,\, \left\{\frac{1}{2}h'D_0 h'+h'\partial_{\bot}^{2} h + \nu_{\parallel 0}\,
h'\partial_{\parallel}^{2} h -
h'\partial_{\parallel}h^{2}/2\right\}.
\label{acts}
\end{equation}
The coefficient $\nu_{\bot 0}$ was scaled out. 
The bare propagators in the frequency--momentum representation have the following form:
\begin{eqnarray}
\langle hh' \rangle_{0} = \langle h'h
\rangle_{0}^{*}
= \left\{-i \omega+ \nu_{\parallel 0} k_{\parallel}^2 + k_{\bot}^2
\right\}^{-1}, \nonumber\\
\langle hh \rangle_{0} = 2\pi \delta(\omega)D_0 /
\left\{\nu_{\parallel 0} k_{\parallel}^2 +  k_{\bot}^2
\right\}^2
\label{prop2}
\end{eqnarray}

Anisotropic models have two independent momentum scales $L_{\bot}$ and $L_{\parallel}$
related to the directions perpendicular and parallel to the vector ${\bf n}$. Thus, there are two independent momentum canonical dimensions
$d_{F}^{\bot}$ and $d_{F}^{\parallel}$ such that
\[ [F] \sim [L_{\bot}]^{-d_{F}^{\bot}}
[L_{\parallel}]^{-d_{F}^{\parallel}}.\] The additional normalization conditions are $d_{k_{\bot}}^{\bot}= -d_{\bf x_{\bot}}^{\bot}=1$,
$d_{k_{\bot}}^{\parallel}=-d_{\bf x_{\bot}}^{\parallel}=0$,
$d_{k_{\bot}}^{\omega} = d_{k_{\parallel}}^{\omega}=0$.
The original momentum dimension can be found from the
relation $d_{F}^{k} = d_{F}^{\bot}+ d_{F}^{\parallel}$ (see, e.g., \cite{Book3}).

The canonical dimensions of the fields and the parameters of the theory (\ref{acts}) are summarized in the
table~\ref{tab35}; it is apparent from the the
table~\ref{tab35} that the upper critical dimension is $d_{*}=6$.

\begin{table}
  \begin{center}
   \small
    \begin{tabular}{|c|c|c|c|c|c|c|c|c|}
     \hline
     $F$ & $h'$ & $h$ & $\nu_{\parallel 0}$ & $D_{0}$ & $g_{0}$ & $g$ & $\mu$\\
     \hline

$d_{F}^{\parallel}$ & $2$ & $-1$ & $-2$ & $-3$ & $0$ & $0$&  $0$\\
$d_{F}^{\bot}$ & $d-5$ & $2$ & $2$ & $9-d$ & 
$6-d$ & $0$ & $1$\\
\hline
$d_{F}$ & $d-3$ & $1$ & $0$ & $6-d$ 
& $6-d$ & $0$ & $1$\\
     \hline
    \end{tabular}
  \end{center}
    \caption{\label{tab35}Canonical dimensions of the fields and the parameters of the theory (\ref{acts}).}
\end{table}

The coupling constant $g_0$ and its renormalized counterpart are  defined  by the relations:
\begin{equation}
D_{0}=g_{0} \nu_{\parallel 0}^{3/2}, \quad  D=Z_{g}Z_{\nu_{\parallel}}^{3/2}g\nu_{\parallel}^{3/2}\mu^{\varepsilon},
\end{equation}
where $\varepsilon=6-d$ and $Z_i$ are renormalization constants. The real index of UV divergence (\ref{acts}) is as follows:
\begin{equation}
\delta_{\Gamma}'= 6 - 4N_{h'} -N_{h}.
\label{IndeX22}
\end{equation}
Analysis of counterterms shows that the theory (\ref{acts}) is multiplicatively renormalizable with the renormalized action:
\begin{equation}
S_{R} (H)=\int d\boldsymbol{x} \,\, \left\{\frac{1}{2}h'Dh'+h' 
\partial_{\bot}^{2} h + Z_{1}\nu_{\parallel}h'\,
\partial_{\parallel}^{2} h -Z_{2}h'
\partial_{\parallel}h^{2}/2\right\}.
\label{RenActs}
\end{equation}
Renormalization constants are related as follows:
\begin{equation}
Z_{g}=Z_{1}^{-3/2}Z_{2}^{2},\quad Z_{\nu_{\parallel}}=Z_{1},\quad Z_{h}=Z_{h'}^{-1}=Z_{2}
\label{Zg2}
\end{equation}
and have the following values in one-loop approximation:
\begin{equation}
Z_{1}= 1- \frac{{2\tilde g}}{3\varepsilon}, \quad Z_{2}= 1+ \frac{{\tilde g}}{6\varepsilon},
\end{equation}
where ${\tilde g}=S_{d}/(2\pi)^d g$. For the $\beta$-function it gives:
\begin{eqnarray}
\beta_{g}={\tilde g}(-\varepsilon+4{\tilde g}/3).
\end{eqnarray}

There are two fixed points:
\begin{enumerate}
\item the Gaussian (free) fixed point:
\begin{eqnarray}
g^{*}=0, \quad \partial\beta/\partial g|_{g^{*}}= - \varepsilon.
\end{eqnarray}
\item the fixed point:
\begin{eqnarray}
g^{*}=3\varepsilon/4, \quad \partial\beta/\partial g|_{g^{*}}= \varepsilon.
\label{IRatrs}
\end{eqnarray}
\end{enumerate}
The fixed point (\ref{IRatrs}) is IR attractive at the physically interesting value of spatial dimension $d=2$.

The critical dimension $\Delta_{F}$ of an IR relevant
quantity $F$ in dynamical models with two momentum scales is given by the relation \cite{Book3}:
\begin{eqnarray}
\Delta_{F} = d^{\bot}_{F}+ d^{\parallel}_{F} \Delta_{\parallel} + \gamma_{F}^{*},
\label{dim5}
\end{eqnarray}
where
\begin{eqnarray}
 \Delta_{\parallel}= 1 +\gamma_{\nu_{\parallel}}^{*}/2
\label{dim225}
\end{eqnarray}
and $\Delta_{k_{\bot}} = 1$. Thus, the critical dimensions in one-loop approximation are
\begin{eqnarray}
\Delta_{h} = 1, \quad \Delta_{h'} = d-3, \quad \Delta_{\parallel}=1
\label{dimD3}
\end{eqnarray}
for the Gaussian fixed point and
\begin{eqnarray}
\Delta_{h} = 1-3\varepsilon/8, \quad \Delta_{h'} = 3-3\varepsilon/8,  \quad \Delta_{\parallel}=1+\varepsilon/4
\label{dimD23}
\end{eqnarray}
for the IR-attractive fixed point (\ref{IRatrs}). 

For the pair correlation function of the field $h$ this gives:
\begin{equation}
\langle h({\bf x})\, h({\bf 0}) \rangle \simeq
r_{\bot}^{-2\Delta_{h}}\, {\cal F} \left(
r_{\parallel}/ r_{\bot}^{\Delta_{\parallel}} \right),
\label{dimS}
\end{equation}
where $r_{\bot}=|{\bf x}_{\bot}|$, $r_{\parallel}=x_{\parallel}$ and
${\cal F}$ is a certain scaling function  of critically dimensionless
arguments.

In this case, replacing of the white noise by the spatially quenched one does not change the critical behaviour of the model qualitatively but it shifts the value of the upper critical dimension from $d_{*}=4$ to $d_{*}=6$. Thus, the 
expansion parameter $\varepsilon$ becomes rather large. However, the same upper critical dimension is encountered in models with
cubic interaction: the plain $\Phi^3$ model, the Potts model, models of percolation and random resistor networks (see, e.g., \cite{Book3,P1,P2,P3,P4}).

\section{Static approach to RG analysis of the Pastor-Satorras-Rothman model} \label{sec:Erosion}

The Pastor-Satorras-Rothman model  \cite{Pastor1,Pastor2} is a semiphenomenological model of landscape erosion with a fixed mean tilt. Unlike existing isotropic continuum models of erosion \cite{E8}--\cite{E10} based on diffusion equation with added noise, this model is anisotropic and nonlinear. However, the model in not renormalizable and should be augmented to include an infinite number of coupling constants \cite{US}:
\begin{equation}
\partial_{t} h= \nu_{\bot 0}\, \partial_{\bot}^{2} h + \nu_{\parallel 0}\,
\partial_{\parallel}^{2} h +
\partial_{\parallel}^{2} V(h) + g.
\label{eqe}
\end{equation}
Here $\nu_{\bot 0},\nu_{\parallel 0}$ are viscosity coefficients,  $V(h)$ is a Taylor expansion (\ref{Vh4}), and $g(x)$ is chosen to be the spatially quenched noise (\ref{forceD2}). The problem (\ref{eqe})  is equivalent to the field theory
of the set of fields $H=\{h,h'\}$ with action functional :
\begin{equation}
S(H)=\int d{\bf x} \,\, \left\{h'h'+h'\partial_{\bot}^{2} h + \nu_{\parallel 0}h'\,
\partial_{\parallel}^{2} h +h'
\partial_{\parallel}^{2}\sum_{n=2}^{\infty}\frac{\lambda_{n0} h^{n}}{n!}\right\}.
\label{acte}
\end{equation}
Here we set $D_0=2$ and $\nu_{\bot 0}=1$.

 \begin{table}[h]
  \begin{center}
   \small
    \begin{tabular}{|c|c|c|c|c|c|c|c|c|}
     \hline
     $F$ & $h'$ & $h$ &   $\nu_{\parallel 0}$ & $\lambda_{n0}$ & $g_{n0}$ & $g_n$ & $\mu$\\
     \hline
   $d_{F}^{\parallel}$ & $1/2$ & $1/2$  & $-2$ & $-(n+3)/2$ & $0$ & $0$ & $0$ \\
$d_{F}^{\bot}$ & $(d-1)/2$ & $(d-5)/2$  & $2$ & $d(1-n)/2+(5n-1)/2$ & $(1-n)(d-4)/2$ & $0$ & $1$ \\
\hline
$d_{F}$ & $d/2$ & $d/2-2$ & $0$  & $(1-n)(d-4)/2$ & $(1-n)(d-4)/2$ & $0$ & $1$\\
     \hline
    \end{tabular}

  \end{center}
  \caption{\label{tab4}Canonical dimensions of the fields and the parameters in the
model (\ref{acte}).}
\end{table}

Canonical dimensions of the fields and the parameters of the theory (\ref{acte}) are summarized in the table~\ref{tab4}. The theory is multiplicatively renormalizable with upper critical dimension $d_{*}=4$ and renormalized action: 
\begin{equation}
S_{R} (H)=\int d\boldsymbol{x} \,\, \left\{h'h'+ h'
\partial_{\bot}^{2} h + Z_{\parallel}\nu_{\parallel}h'\,
\partial_{\parallel}^{2} h +
h'\partial_{\parallel}^{2}\sum_{n=2}^{\infty}\frac{Z_n\lambda_{n} h^{n}}{n!}\right\}.
\label{RenActe}
\end{equation}
The bare coupling constants  $g_{n0}$ and their renormalized counterparts $g_{n}$ are related with parameters
$\lambda_{n0}$ and $\lambda_n$ as follows:
\begin{equation}
\lambda_{n0}=g_{n0} \nu_{\parallel 0}^{(n+3)/4}, \quad
\lambda_{n}=g_{n} \nu_{\parallel}^{(n+3)/4}\mu^{\varepsilon(n-1)/2},
\label{I6}
\end{equation}
where $\varepsilon=4-d$.  Divergent part of the one-loop term $\Gamma^{(1)}(H)$ of the generating functional $\Gamma(H)$ is:
\begin{equation}
\Gamma^{(1)}(H)= \frac{S_d}{(2\pi)^d}\frac{\mu^{-\varepsilon}}{ \varepsilon}\int dx
\frac{V''(h(x))}{\sqrt{(\nu_{\parallel}+V'(h(x)))}}\, \partial^{2}h'(x).
\label{I118}
\end{equation}
This gives the following one-loop expressions for the renormalization constants:  
\begin{equation}
Z_{\parallel}=1-\frac{r_{1}S_d}{(2\pi)^d\varepsilon}, \quad
Z_{n}=1-\frac{r_{n}}{g_{n}}\frac{S_d}{(2\pi)^d\varepsilon},
\label{I201}
\end{equation}
where the first four parameters $r_n$ are:

\[ r_{1}=g_{3}-\frac{1}{2}g_{2}^{2},\quad r_{2}=g_{4}-\frac{3}{2}g_{2}g_{3}+\frac{3}{4}g_{2}^{3}, \]
\[ r_{3}=g_{5}-2g_{2}g_{4} -\frac{3}{2}g_{3}^{2}+\frac{9}{2}g_{2}^{2}g_{3}-\frac{15}{8}g_{2}^{4}, \]
\[  r_{4}=g_{6}-\frac{5}{2}g_{2}g_{5} +\frac{15}{2} g_{2}^{2} g_{4} -5 g_{3}g_{4}+
\frac{45}{4} g_{2}g_{3}^{2}-\frac{75}{4} g_{2}^{3}g_{3} +\frac{105}{16} g_{2}^{5}. \]

This gives for RG functions: 
\begin{eqnarray}
\gamma_{\parallel} &=& a (2g_{3}-g_{2}^{2}),
\label{I23a} \\
\beta_{2}&=&-\frac{\varepsilon}{2} g_{2}+a(-g_{4} +\frac{11}{2} g_{2}g_{3}-\frac{11}{4} g_{2}^{3}),
\nonumber \\
\beta_{3}&=&-\varepsilon g_{3}+a(-2g_{5}+4g_{2}g_{4}+6 g_{3}^{2}-\frac{21}{2} 
g_{2}^{2}g_{3} +\frac{15}{4}  g_{2}^{4}),
\nonumber \\
\beta_{4}&=&-\frac{3\varepsilon}{2} g_{4}+a(-2g_{6}+5g_{2}g_{5}+\frac{27}{2}g_{4}g_{3}-\frac{67}{4}g_{4}g_{2}^{2}-\frac{45}{2}g_{3}^{2}g_{2}+\nonumber \\
&&\frac{75}{2}g_{2}^{3}g_{3}-\frac{105}{8}g_{2}^{5}),
\label{I23b}
\end{eqnarray}
where $a\equiv 2S_d/2(2\pi)^d$.

The RG equations have a two-dimensional surface of fixed points parametrized by the values of $g_{2*}$, and $g_{3*}$
in the infinite-dimensional space of the 
couplings $g\equiv\{g_{n}\}$. If the surface contains IR attractive regions then the system exhibits scaling behavior with non-universal critical exponents that satisfy the exact relation
$2\Delta_{h}=d-5+\Delta_{\parallel}$. One loop approximation gives $\Delta_{\parallel}=1+a (2g_{3*}-g_{2*}^{2})/2$, $\Delta_{h}=a (2g_{3*}-g_{2*}^{2})/4$. 

Let us recall that in the case of the white noise  the exact relation was $2\Delta_{h}=d-3+\Delta_{\parallel}$ \cite{US}. In the notations used in \cite{Delamotte} ($\alpha=-\Delta_{h}$, $\zeta=\Delta_{\parallel}$) these expressions become 
\begin{eqnarray}
-2\alpha&=&d-5+\zeta \label{1}
\end{eqnarray}
for the quenched noise and
\begin{eqnarray}
-2\alpha&=&d-3+\zeta \label{2}
\end{eqnarray}
for the white noise. Let us consider the case of $d=2$ and assume that $\alpha\ge0$ and $\zeta\ge1$ (these inequalities arise from the experimental data \cite{Pastor1,Pastor2,Delamotte}). Then the inequalities give only the trivial solution for the expression (\ref{2}) (the white noise):  $\alpha=0$ and  $\zeta=1$. However, they give the range of values for the both exponents for the expression (\ref{1}) (the quenched noise): $0\le\alpha\le1$ and  $1\le\zeta\le3$. This result is not surprising if we remember that the upper critical dimension for the white noise is $d_{*}=2$. Indeed, the model would only predict trivial scaling in its upper critical dimension, the fact that was confirmed in \cite{Delamotte}. The spatially quenched disorder, however, yields upper critical dimension $d_{*}=4$. Thus, non-trivial scaling behavior is possible in $d=2$. In \cite{Delamotte} it was shown that there is indeed a line of IR attractive fixed points in $d=2$ that corresponds to the critical exponent $0\le\alpha\le1$. Though, it is not clear whether this line of the fixed point can be reached from the physical initial conditions (it could be unphysical like in KPZ model). Lastly, it is also not clear whether the spatially quenched noise caused the non-universality or whether it only made the scaling in $d=2$ non-trivial.

\section{Conclusion} \label{sec:Conclusion}

In this paper, we apply the field theoretic renormalization group to several models of nonlinear dynamics of various fluctuating surfaces. Those models are described by stochastic differential equations of the type (\ref{problem}) subjected to a random noise. The latter is taken to be Gaussian with the time-independent correlation function of the form (\ref{forceD2}).

In contrast to more conventional models with white in-time noise with the correlator (\ref{forceD}), such models allow for two different field theoretic formulations.

One possibility, adopted, e.g., in \cite{Delamotte,Vestnik}, is to derive in a straightforward fashion the dynamic De Dominicis—Janssen action functional (\ref{act}) and then to apply the dynamic form of the field theoretic RG (in earlier works, e.g., \cite{Pastor1,Pastor2}, the dynamic Wilsonian RG was used).
Owing to the presence in the action functional (\ref{act}) of the nonlocal in-time contribution (\ref{1t2}), the standard dimensional analysis of the UV divergences in this approach should be somewhat modified, see discussion in Section \ref{sec:Static}.

In this paper, we proposed and applied to four different models an alternative approach. From the very beginning, it deals with time-independent fields, involves simpler practical calculations and requires only standard analysis of divergences.

The main results are as follows.

The KPZ equation with the static noise appears logarithmic at $d=4$ (i.e., $d_{*}=4$ is its upper critical dimension), which, on principle, allows for a power-like scaling behavior in the most interesting dimensions $d=2$ and 3. The corresponding RG equations have a nontrivial fixed point, but it is either IR repulsive or lies in the unphysical region of the parameters. The model shares this drawback with its more conventional dynamic version, where it is sometimes referred to as “a major open problem in nonequilibrium statistical physics” \cite{LK}. For that model, existence of hypothetical strong-coupling fixed point, not accessible for any kind of perturbative analysis, is supported by the functional RG \cite{Canet,Canet2}. It would be interesting to apply the functional approach to the static version of the KPZ model.

For the Hwa-Kardar model of self-organized criticality, existence of an appropriate fixed point was established, and the corresponding exponents were calculated in the leading one-loop approximation.

It was shown that the RG analysis of Pastor-Satorras--Rothman model of landscape erosion and Pavlik’s ramification of the KPZ model necessarily involve infinitely many interaction vertices and, hence, corresponding coupling constants. In this respect, those models are also similar to their dynamic counterparts with the noise (\ref{forceD}). In all those cases, the RG equations possess one- or two-dimensional manifolds of fixed points in the infinite-dimensional space of couplings. It seems very likely that regions of IR attractiveness exist on those curves or surfaces. If so, scaling behavior with non-universal critical exponents will be observed.

The results obtained are in agreement with those derived using the standard (dynamic) approach to the four models in \cite{Vestnik}. They are also consistent with the results obtained for the Pastor-Satorras--Rothman model within the functional RG \cite{Delamotte}. 

Let us conclude with a brief comparison of the two approaches. The standard approach with the dynamic action (\ref{act}) is unavoidable when the effects of time-dependent disturbances are studied, like, e.g., the interaction with a randomly moving medium (e.g., in \cite{Us,AK1}). On the other hand, presented approach with the static action (\ref{act2}) seems to be very promising for the application of the
non-perturbative functional RG: derivation of the exact functional RG equations and construction of the appropriate ansatz for the effective action will definitely be simpler if the time variable is excluded from the very beginning. Most interesting here is probably the search for the strong- coupling non-perturbative fixed point in the static version of the KPZ model. This work is in progress.

\section*{Acknowledgments}

We are thankful to C. Duclut for bringing the work \cite{Delamotte} to our attention. We also thank L. Ts. Adzhemyan, N. M. Gulitskiy, M. Hnatich, M. V. Kompaniets, and M. Yu. Nalimov for fruitful discussions. The work was supported by the Foundation for the Advancement of Theoretical Physics and Mathematics ``BASIS''.

 \end{document}